\documentstyle[prl,twocolumn,aps]{revtex}
\input{epsf}
\begin{document}
\draft
\twocolumn[\hsize\textwidth\columnwidth\hsize\csname @twocolumnfalse\endcsname
\title{Transition to an Insulating Phase Induced by Attractive Interactions \\
in the Disordered Three-Dimensional Hubbard Model}
\author{Bhargavi Srinivasan$^{(a)}$, Giuliano Benenti$^{(a,b)}$, 
and Dima L. Shepelyansky$^{(a)}$} 
\address{$^{(a)}$Laboratoire de Physique Quantique, UMR 5626 du CNRS,
Universit\'e Paul Sabatier, 31062 Toulouse Cedex 4, France}
\address{$^{(b)}$International Center for the Study of Dynamical
Systems, Universit\`a degli Studi dell'Insubria and}
\address{Istituto Nazionale per la Fisica della Materia,
Unit\`a di Como, Via Valleggio 11, 22100 Como, Italy}  
\date{March 25, 2002}
\maketitle

\begin{abstract}
We study numerically the interplay of disorder and attractive interactions
for spin-1/2 fermions in the three-dimensional Hubbard model. The
results obtained by projector quantum Monte Carlo simulations show that
at moderate disorder, increasing the attractive interaction leads to 
a transition from delocalized superconducting 
states to the insulating phase of
localized pairs. This transition takes place well
within the metallic phase of the single-particle 
Anderson model.
\end{abstract} 
\pacs{PACS numbers: 71.30.+h, 74.20.-z, 71.10.-w}
\vskip1pc]

\narrowtext

Two limiting cases of the non-trivial problem of
quantum transport in three dimensions (3D) in the
presence of disorder and attractive interactions 
between particles  were worked out by 
Anderson in the late 1950s \cite{anderson0,anderson}. 
In the limit of weak interactions, the increase of
disorder leads to the Anderson transition from 
metallic transport to the localized insulating phase \cite{anderson0}.
Whereas, in the absence of disorder, attractive interactions
between spin-1/2 fermions lead to the appearance
of BCS superconductivity which is not affected by the
introduction of weak disorder \cite{anderson}.
These limiting cases have been extensively investigated
and detailed information is now available for the
one-particle Anderson transition (see e.g. \cite{tvr,kramer0,mirlin}) 
and the weakly disordered
BCS superconductor (see e.g. \cite{degennes,larkin,ralph}).
However, a theoretical treatment of the intermediate 
regime, where both disorder and interactions are important,
is difficult due to the absence of relevant small
parameters. New results on the physical properties of transport
in this regime are therefore of fundamental interest. Furthermore, 
an understanding of this realistic regime
would contribute to the interpretation of recent experiments 
on 3D superconductors,
where both disorder and
interactions are naturally present in the physical systems studied 
\cite{boebinger,ando,gantmakher,samoilov}. Indeed, the experimental results
of Ref. 
\cite{boebinger} indicate an intriguing correlation between the
Anderson transition in a strong magnetic field and optimal doping
for the superconducting transition temperature. Also,
an explanation of the unusual resistivity dependence on magnetic field observed
in Ref. \cite{gantmakher} requires 
a better understanding  of the interplay
between disorder and attractive interactions.
In addition, recent breakthroughs in cold-atom experimental techniques 
have provided new possibilities for investigations 
of interacting atoms on 3D optical lattices, leading to the  
observation of a superfluid to Mott insulator 
quantum phase transition for ultra-cold atoms \cite{greiner}.
These extremely clean experiments open unprecedented possibilities
for precise studies of lattice models with experimentally tunable 
interactions and provide new challenges for theoretical investigations.

Numerical simulations provide a valuable tool for the study of the
non-trivial regime where both interactions and disorder play a relevant role.
Among various numerical approaches, quantum Monte Carlo methods
constitute the most promising possibility for the 
simulation of 3D systems with a large number of particles 
\cite{scalettar,imada,ceperley,gubernatis}. 
These methods have several advantages compared to 
exact-diagonalization approaches which are restricted to a relatively small
number of particles \cite{didier,pichard0}. 
In this work we use the projector quantum Monte Carlo method (PQMC) to 
perform  numerical simulations of the
3D Anderson transition in the presence of attractive interactions.
To the best of our knowledge, this approach allows us to explore this
new regime for the first time.

To investigate the interplay of disorder and attractive interactions, 
we study numerically the disordered 3D Hubbard model  
with $N$ fermions on a cubic lattice with $L^3$ sites. 
The Hamiltonian is defined by
\begin{equation}
\label{hamiltonian}
H=-t\sum_{<{\bf ij}>\sigma}
c_{{\bf i}\sigma}^\dagger c_{{\bf j}\sigma}
+\sum_{{\bf i}\sigma} \epsilon_{\bf i}
n_{{\bf i}\sigma} +U \sum_{\bf i} n_{{\bf i} \uparrow}
n_{{\bf i} \downarrow},
\end{equation}
where $c_{{\bf i}\sigma}^\dagger$ ($c_{{\bf i}\sigma}$)
creates (destroys) a spin-1/2 fermion at site ${\bf i}=(i_x,i_y,i_z)$ 
with spin $\sigma$,
$n_{{\bf i}\sigma}=c_{{\bf i}\sigma}^{\dagger}
c_{{\bf i}\sigma}$ is the corresponding
occupation number operator.    
The hopping term $t$ between nearest neighbor lattice sites
parameterizes the kinetic energy and
the random site energies $\epsilon_{\bf i}$ are homogeneously 
distributed in the interval 
 $[-W/2,W/2]$, where $W$ determines the disorder strength. The parameter
$U$ measures the strength of the screened attractive 
Hubbard interaction ($U<0$)
and periodic boundary conditions are taken in all directions. 
At $U = 0$, 
the Hamiltonian (\ref{hamiltonian})  reduces to the one-body 
Anderson model, which exhibits  a metal-insulator transition in 
three dimensions \cite{tvr,kramer0,mirlin}. 
For $W=0$ the Hamiltonian corresponds to the clean attractive Hubbard model, 
with a superconducting ground state in 3D. 

We study this model by the PQMC method, which is 
an efficient method for the investigation of the
ground state properties of interacting fermion systems 
\cite{imada,bhargavi}.
For attractive Hubbard interactions ($ U < 0$), 
there is no sign problem and
the method is exact up to
discrete time-step and statistical errors, which can be well-controlled.
We consider $N=14,32,62$, and $108$ particles on a cubic
lattice of size $L=3,4,5$, and $6$, respectively, at 
an approximately constant filling factor 
$\nu=N/(2 L^2)\approx 1/4$, $2\leq W/t \leq 10$, and 
$U/t=-4$. For each disorder realization,
we used a discrete  Trotter 
decomposition  with a time step $\Delta\tau=0.1$  and 
projected through $60$ time steps. 
In total we carried out
$3000$ Monte Carlo sweeps for each simulation, with  approximately 1000  
sweeps for equilibration. With these PQMC parameters 
we obtained good convergence of the computed physical quantities.  
The results  are averaged over $N_R\geq 12$ disorder 
realizations, except for the most time-consuming simulations 
at $L=6$, where $N_R=6$. The simulations are carried out 
in the sector with the total spin component $S_z=0$. 
This is sufficient to study the 
ground state properties, since all of the eigenvalues of 
the Hamiltonian (\ref{hamiltonian}) belong to the 
spectrum of the $S_z=0$ subspace.

A quantitative measure of the localization properties of 
the system can be obtained, even in the presence of interactions. 
It is based on  the
probability density distribution for an added pair at the
Fermi edge.
This distribution is approximately equal to the
charge density difference 
$\delta\rho({\bf i})=\rho({\bf i},N+2)-\rho({\bf i},N)$, 
where $\rho({\bf i})$ is the ground state charge density 
at site ${\bf i}$ ($\sum_{\bf i}\delta\rho({\bf i})=2$).
The values of $\rho({\bf i},N)$, $\rho({\bf i},N+2)$ 
are obtained from two independent PQMC simulations for the
same disorder realization.
For $U=0$, this difference  is identical  to the 
single-particle probability distribution for the  
eigenstate $\psi_F({\bf i})$ of the Anderson model at the Fermi 
level, with $\delta\rho({\bf i})/2=|\psi_F({\bf i})|^2$. 
From this distribution, we obtain   
the inverse participation ratio (IPR) 
for an added pair,
$\xi=(\sum_{\bf i}(\delta\rho({\bf i})/2)^2)^{-1}$, 
which determines the number 
of sites visited by this pair. At $U=0$, this 
quantity reduces to the usual one-particle IPR at the Fermi level. 
For $U\neq 0$, since the interaction is screened and short-ranged,
$\xi$ still determines the localization properties
of pairs in the vicinity of the Fermi level.

\begin{figure} 
\vspace{-0.9cm}
\centerline{\epsfxsize=4.2cm\epsffile{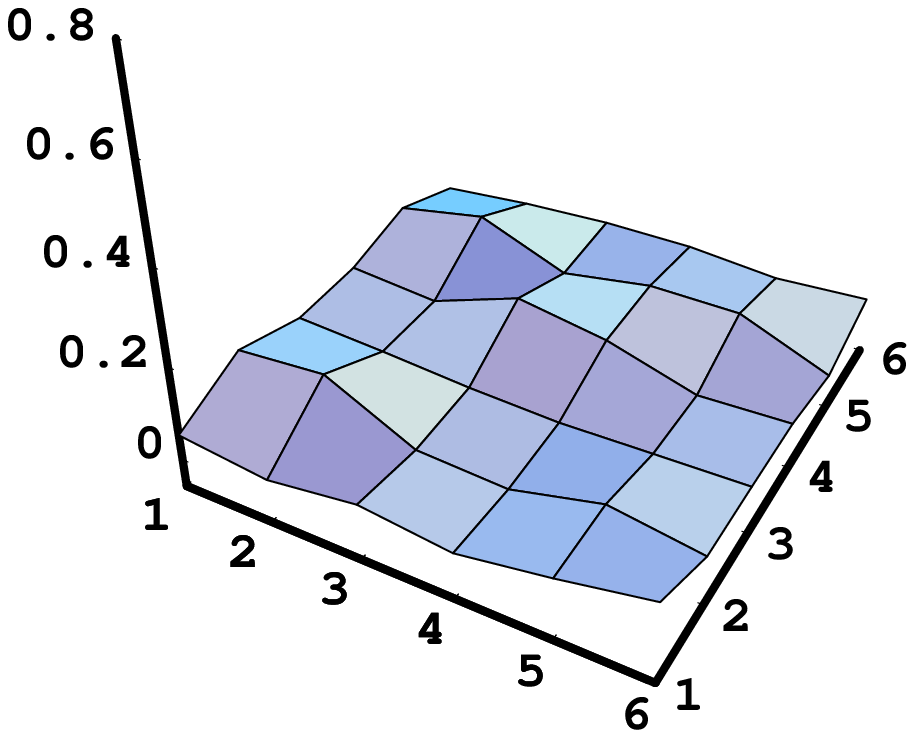}
\hfill\epsfxsize=4.2cm\epsffile{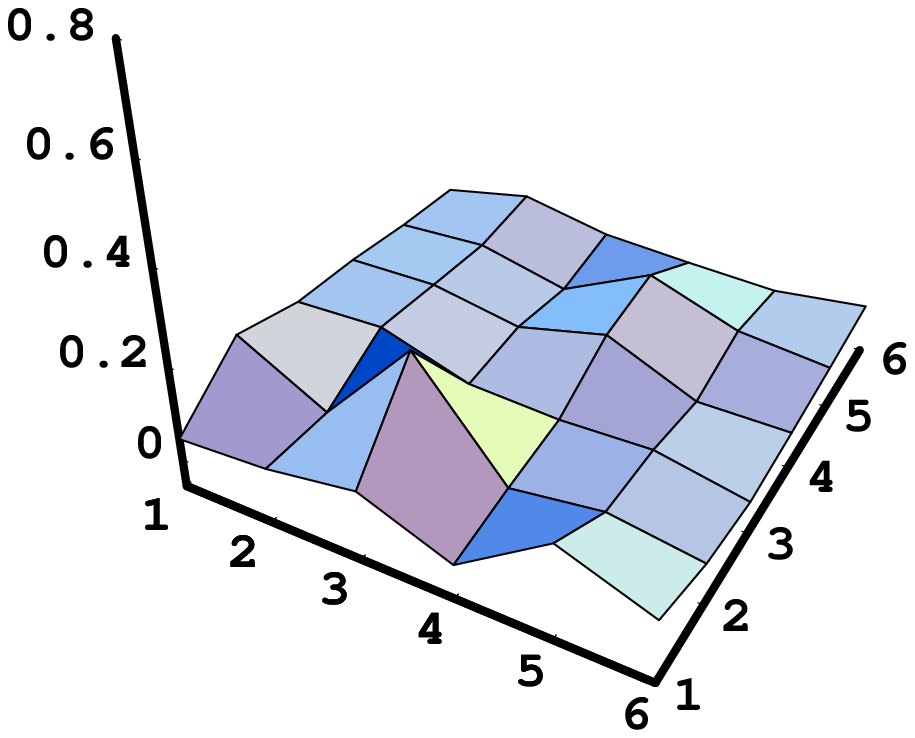}}  
\vspace{-.2cm}
\centerline{\epsfxsize=4.2cm\epsffile{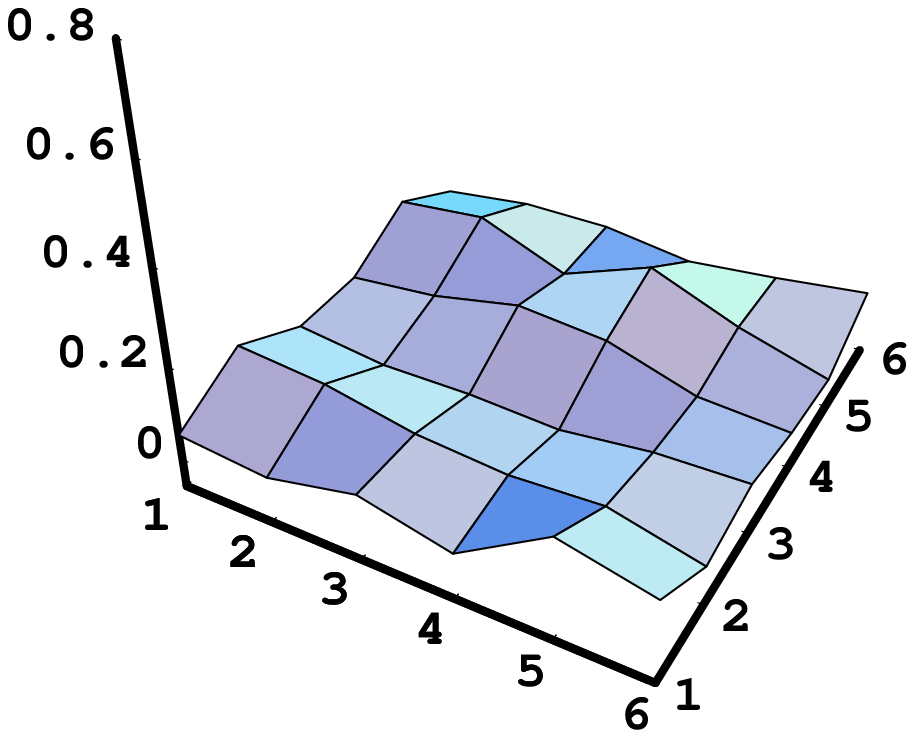}
\hfill\epsfxsize=4.2cm\epsffile{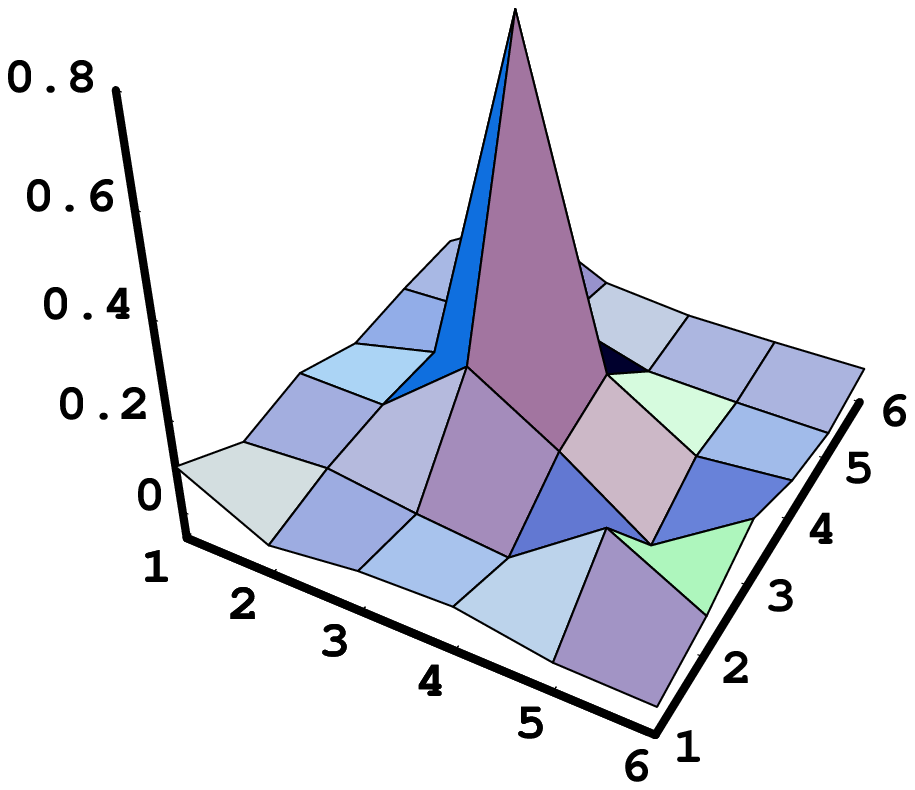}}  
\vspace{-0.2cm}
\centerline{\epsfxsize=4.2cm\epsffile{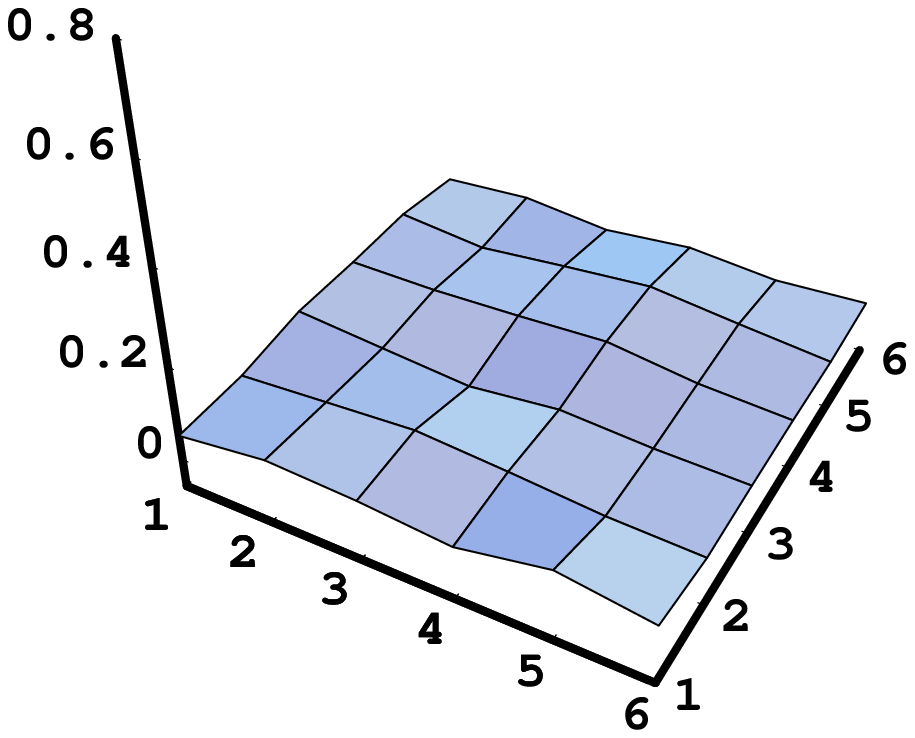}
\hfill\epsfxsize=4.2cm\epsffile{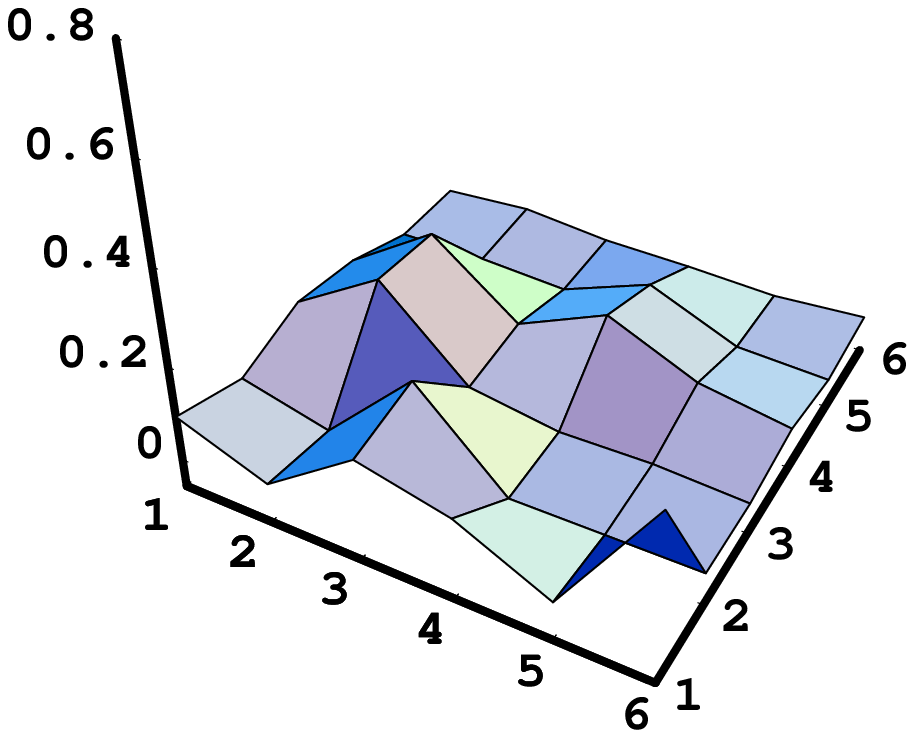}}  
\caption{Distribution of charge  density difference for an added
pair, $\delta\rho_p$, projected on 
the $(x,y)$ plane for a $6\times 6 \times 6$ lattice 
for the same single disorder realization, 
with $W/t=2$ (left) and $W/t=7$ (right), $N=108$.
Top:  exact computation for $U=0$, $\xi=70; 55$ (left; right).
Middle: PQMC calculation for  $U/t=-4$, $\xi=48; 6.5$ (left; right).
Bottom: BdG mean field calculation 
for $U/t=-4$, $\xi=132; 25$ (left; right).} 
\label{fig1} 
\end{figure}

A typical example of $\delta\rho({\bf i})$ for an
added pair  
is shown in Fig.\ref{fig1} for 
a single disorder realization. 
For graphical representation $\delta\rho({\bf i})$ is projected 
on the $(x,y)$ plane, giving 
$\delta\rho_{p}(i_x,i_y)=\sum_{i_z} \delta\rho(i_x,i_y,i_z)$.  
The  plots of Fig.\ref{fig1} (left) refer to the weakly disordered 
regime ($W/t=2$). They clearly show delocalization of the added pair,
 both for free particles (top, $U=0$) and 
for the attractive Hubbard model (middle, $U/t=-4$). 
It should be noted that this disorder strength is much smaller 
than the critical disorder strength of the 3D Anderson transition
which takes place at $W<W_c(U=0)\approx 16.5 t$ \cite{zhar}.
Therefore, these results confirm Anderson's theorem \cite{anderson} 
according to which the Cooper pairs remain delocalized at 
weak disorder.
Thus, the superconducting 
phase is not affected by weak disorder, since Cooper pairs can 
be formed by pairing the time-reversed eigenstates of the 
corresponding non-interacting disordered problem. 

A qualitatively new situation appears at a stronger disorder 
strength $W/t=7$ (Fig.\ref{fig1} (right)). 
Here, the two added particles are delocalized at $U=0$,
since we are still 
inside the metallic single particle phase $W<W_c(U=0)$.
On the contrary, a pronounced peak appears for $\delta\rho({\bf i})$ at 
$U/t=-4$  clearly showing the formation of  localized pairs.
This is borne out by the IPR which 
 drops from $\xi=55$ at $U=0$ to $\xi=6.5$ at 
$U/t=-4$. This effect gives an indication 
that attractive interactions induce localization in the 
metallic regime of the non-interacting model, leading to the formation
of bi-particle localized states. 

The results of the Bogoliubov-de Gennes (BdG) mean field 
calculation \cite{degennes,trivedi}, 
are shown in the two bottom plots of Fig.\ref{fig1}.
We note that, within BdG approach, at weak disorder strengths
($W/t = 2$), interactions 
smooth out charge fluctuations leading to an increase of
the IPR, from  
$\xi=70$ at $U=0$ to $\xi=132$ at $U/t=-4$. On the contrary, 
at stronger disorder strengths ($W/t = 7$) interactions slightly
favor localization even 
within BdG approximation where the IPR drops from 
$55$ to $25$ when $U/t$ goes from $0$ to $-4$. 
This happens because the mean 
field treatment of interactions introduces a site dependent 
Hartree shift $U_H({\bf i})=|U|\rho({\bf i})/2$ \cite{trivedi}. 
At strong disorder, when the charge density $\rho({\bf i})$ 
is highly inhomogeneous, this term acts as an additional 
disorder potential \cite{note}. 
However, the main problem with the BdG approach is that
many local minima appear and convergence becomes problematic even 
at moderate disorder strengths and
system sizes ($W/t > 7, L \geq 6$).
Furthermore, it is clear from the comparison with the  
PQMC charge density difference at $W/t=7$ that important effects 
beyond mean field cannot be reproduced within BdG approach.

\begin{figure} 
\centerline{\epsfxsize=8.cm\epsffile{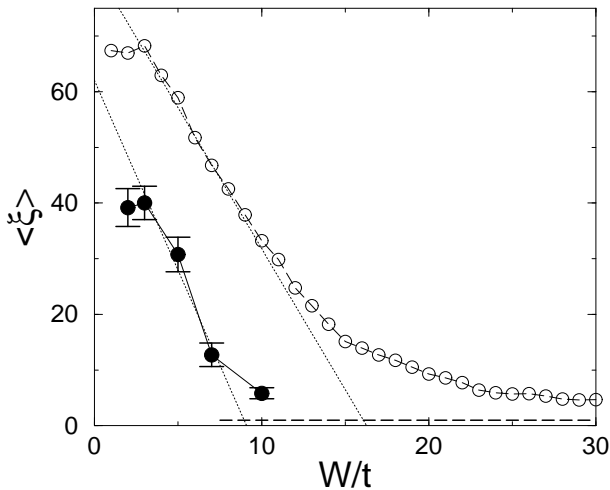}} 
\caption{Inverse participation ratio $<\xi>$ averaged over
disorder realizations, as 
a function of disorder strength $W$ for a 
$6\times 6\times 6$ lattice, at $U=0$ (empty 
circles) and $U/t=-4$ (filled circles).  
Dotted lines show  linear fits to the data,
the dashed line represents $\xi = 1$ (see text) and
error bars indicate statistical errors. 
} 
\label{fig2} 
\end{figure} 

A more quantitative description of the localization effect 
induced by attractive interactions can be seen from the
dependence of
IPR $<\xi>$ on disorder strength, $W$,  shown in Fig.2,
with $<\xi>$ averaged over realizations of disorder. 
The data clearly show that interactions lead to a 
significant reduction 
of $<\xi>$.  For $U=0$, the dependence of $<\xi>$ on $W$
is characterized by  two distinct regions:
a  relatively flat region for large $W$ where $<\xi>$ 
slowly approaches the asymptotic value of 1 and another
region in which $<\xi>$ grows with decreasing $W$.
At very weak disorder $<\xi>$ remains bounded by the total number
of lattice sites. The linear fit of the $<\xi>$ data in
the second region crosses the $<\xi> = 1$ line at 
$W_c(U=0)\approx 16t$ which is very close to the exact value of $W$
for the Anderson transition for non-interacting particles,
$W_c\approx 16.5t$ \cite{zhar}. A similar analysis  carried
out for the $<\xi>$ data in the presence of interactions
gives transition from delocalized to localized 
pairs at  $W_c(U=-4t)\approx 9t$. 

\begin{figure} 
\centerline{\epsfxsize=8.cm\epsffile{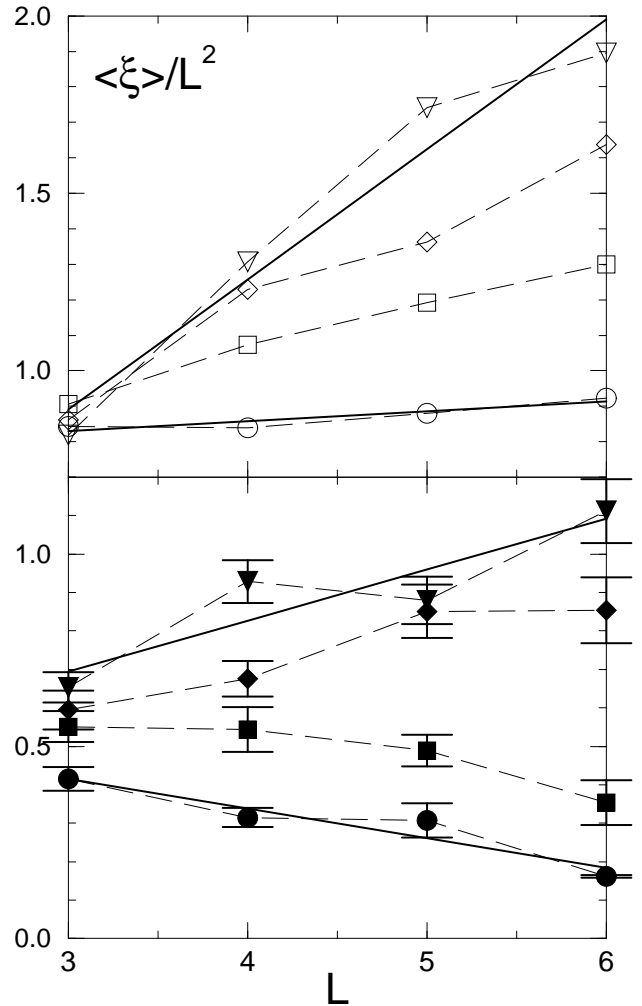}} 
\caption{Dependence of the scaled inverse participation ratio 
$<\xi>/L^2$  on the linear dimension of the system $L$,  
for $U=0$ (top) and $U/t=-4$ (bottom), 
with $W/t=3$ (triangles), $5$ (diamonds), 
$7$ (squares), and $10$ (circles).
Error bars show statistical errors.
The straight line fits show the 
average dependence on $L$ for the extremal values of $W$ studied,
designating the transition from superconducting to insulating
behavior for $U/t = -4$.
}  
\label{fig3} 
\end{figure} 

Further evidence for the interaction induced transition
comes from  a finite-size scaling 
analysis. The relevant dimensionless quantity for such an
analysis is the system conductance $g(L)$ 
\cite{tvr,imry}. For large $g(L)$, the macroscopic transport 
theory in the 3D delocalized phase gives $g(L)\propto L \propto \xi/L^2$ since
$\xi\propto L^3$ for delocalized wave functions. 
For the localized phase, $\xi$ is determined by the localization
length $l$, and is independent of the system size $L$, with
$\xi\sim l^3$. 
Hence, the ratio $\xi/L^2$ falls off as $1/L^2$ in this regime.
Therefore the transition point  $g(L)\approx 1$ 
can be located from the condition 
$\xi(L)/L^2 = \hbox{const}$. 
The results of the finite-size scaling analysis for the scaled ratio
$\xi/L^2$ are shown in 
Fig.\ref{fig3} for $3\leq L \le 6$. The range of disorder values studied
($3t\leq W \leq 10t$) corresponds to the 
metallic side of the single-particle Anderson 
transition. Therefore the scaling 
analysis at $U=0$ shows an increase of $\xi/L^2$ 
with the system size. A strikingly different situation
appears at $U=-4t$: 
at $W/t=3$ and $5$ the scaling 
ratio $\xi/L^2$ still grows with the system size, while
at $W/t=7$ and $10$ this quantity drops with $L$. 
This suggests the appearance of 
a superconductor to insulator transition induced by attractive interactions, 
with the transition  point 
$W_c(U=-4t)\approx 6t$ at the thermodynamic limit. 
This value is in reasonable agreement with the value
obtained for a single system size in Fig. 2.
A precise location of the transition point would require 
a significant increase of the system sizes and a larger
number of disorder realizations.
However the results obtained in the present study clearly
show the transition to an insulating phase at
disorder strengths being less than half the value of the
critical disorder strength for the single-particle
Anderson transition.
Hence, in the presence of attractive interactions,
the insulating phase penetrates 
inside the metallic non-interacting phase. 
This unexpected result can be understood 
on the basis of  the following physical 
argument \cite{jose}. The attractive interaction creates pairs 
of effective mass $m^\star$ twice as large as  the single fermion mass $m$. 
This halves the effective hopping term $t^\star \propto 1/m^\star = 1/2m$ 
which induces an increased  
effective disorder $W/t^\star=2W/t$
and thus enhances localization effects. Such an argument predicts the decrease 
in critical disorder strength by a factor of 2, which is 
in reasonable agreement with our results.

In conclusion we show that in disordered systems, attractive interactions
that lead to superconductivity at weak disorder also lead 
to the insulating phase of localized pairs at moderate disorder strengths, 
well within the metallic phase of
non-interacting fermions. Thus, by increasing the attraction strength,
one can go from  superconducting  to insulating behavior. This is also
possible by increasing disorder.
Of course, experimentally, it is not easy to vary the interaction and disorder strengths.
However, indirectly, this can be achieved by introducing a 
relatively strong magnetic field. 
This magnetic field can increase  the effective disorder strengths
since it forces an electron to return to the impurity \cite{berglund}.
At the same time, it also effectively decreases attraction by 
pair breaking.  Thus, the increase of magnetic field may first 
increase the disorder and drive the system  from superconductor to 
insulator with localized pairs as seen experimentally in \cite{gantmakher}.
Further increase of magnetic field breaks pairs and leads 
to  a transition from insulating phase of localized pairs to 
a metallic phase of almost non-interacting fermions.
Such a scenario leads to a transition from superconductor to insulator,
followed by an insulator to metal transition  with
increasing magnetic field, in qualitative agreement with experimental observations
\cite{gantmakher}.

We thank the IDRIS in Orsay for access to their supercomputers. 
This work was supported in part by the EC  RTN contract HPRN-CT-2000-0156.


\begin{thebibliography}{99}
\bibitem{anderson0} P.W. Anderson, Phys. Rev. {\bf 109}, 1492 (1958).
\bibitem{anderson} P.W. Anderson, J. Phys. Chem. Solids 
                {\bf 11}, 26 (1959). 
\bibitem{tvr} P.A.~Lee and T.V.~Ramakrishnan, Rev. Mod. Phys.
                {\bf 57}, 287 (1985).
\bibitem{kramer0}  A. MacKinnon and B. Kramer, Rep. Prog.
                 Phys. {\bf 56}, 1469 (1993).
\bibitem{mirlin} A. D. Mirlin, Phys. Rep. {\bf 326}, 259 (2000).
\bibitem{degennes} P.G. de Gennes, {\it Superconductivity 
                 of metals and alloys} (Benjamin, New York, 1966).
\bibitem{larkin} A. Larkin, Ann. Phys. (Leipzig), {\bf 8}, 785 (1999).
\bibitem{ralph}  J. von Delft and  D. C. Ralph, Phys. Rep.
                 {\bf  345}, 61 (2001).
\bibitem{boebinger} G. S. Boebinger {\it et al.}, 
                 Phys. Rev. Lett. {\bf 77}, 5417 (1996).
\bibitem{ando} S. Ono {\it et al.}, 
               Phys. Rev. Lett. {\bf 85}, 638 (2000).
\bibitem{gantmakher} V. F. Gantmakher {\it et al.}, 
                  JETP Lett. {\bf 68}, 363 (1998) 
                  [Pis'ma Zh. Eksp. Teor. Fiz. {\bf 68}, 337 (1998)];
                  Ann. Phys. (Leipzig) {\bf 8}, SI-73 (1999)
                  (cond-mat/0004377).
\bibitem{samoilov} A. V. Samoilov, N.-C. Yeh and C. C. Tsuei,
                  Phys. Rev. B {\bf 57}, 1206 (1998).
\bibitem{greiner} M. Greiner {\it et al.}, Nature  {\bf 415}, 39 (2002).
\bibitem{scalettar} S.R. White {\it et al.},
                  Phys. Rev. B {\bf 40}, 506 (1989). 
\bibitem{imada} M. Imada and Y. Hatsugai, J. Phys. Soc. Jpn.
                  {\bf 58}, 3752 (1989).  
\bibitem{ceperley} D. M. Ceperley, Rev. Mod. Phys. {\bf 67}, 279 (1995).
\bibitem{gubernatis} M. Jarrell and J. E. Gubernatis, Phys. Rep.
                  {\bf 269}, 133 (1996).
\bibitem{didier} G.Bouzerar, D.Poilblanc and G.Montambaux, Phys. Rev. B 
                  {\bf 49}, 8258 (1994).
\bibitem{pichard0} J-L. Pichard, G. Katommeris and F. Selva, to appear
                  in {\it Proc. of XXXVIth Rencontres de Moriond}
                  (Eds. T. Martin and G. Montambaux)
                  (cond-mat/0107500).
\bibitem{bhargavi} B. Srinivasan and D.L. Shepelyansky, 
                  Eur. Phys. J. B {\bf 24}, 469 (2001). 
\bibitem{zhar} At quarter filling $W_c\approx 16.5 t$, 
               see Fig.2 in T. Dr\"ose {\it et al.}, 
               Phys. Rev. B {\bf 57}, 37 (1998). 
\bibitem{trivedi} A. Ghosal, M. Randeria, and N. Trivedi, Phys. Rev. Lett. 
         {\bf 81}, 3940 (1998); Phys. Rev. B {\bf 65}, 014501 (2001). 
\bibitem{note} To assess the localizing effect of the 
          Hartree shift, we note that in the case of Fig.\ref{fig1} bottom 
          right one would get complete delocalization ($\xi=151$) if this term 
          is neglected in the BdG equations.      
\bibitem{imry} Y. Imry, {\it Introduction to Mesoscopic Physics}, 
              Oxford Univ. Press (1997).
\bibitem{jose} J. Lages and D.L. Shepelyansky, Phys. Rev. B 
                 {\bf 62}, 8665 (2000). 
\bibitem{berglund} N.~Berglund {\it et al.}, 
             Phys. Rev. Lett. {\bf 77}, 2149 (1996).
\end{thebibliography}
\end{document}